\documentclass[aps,pra,twocolumn,notitlepage,amsmath,amstex,amssymb,longbibliography,floatfix,a4paper,twocolumn,10pt]{revtex4-1}

\usepackage{amsmath,amssymb,graphicx}
\usepackage{framed}
\usepackage{color,times}
\usepackage{xcolor}
\usepackage{afterpage}

\definecolor{darkblue}{rgb}{0, 0, 0.8}

\usepackage{hyperref}
\hypersetup{
    bookmarks=false,         
    unicode=false,          
    pdftoolbar=false,        
    pdfmenubar=true,        
    pdffitwindow=false,     
    pdfstartview={FitH},    
    pdftitle={Quantum Many-Body Scars and Weak Breaking of Ergodicity},    
    pdfauthor={Maksym Serbyn, Dmitry A. Abanin, and Zlatko Papic},     
    pdfsubject={quantum scars},   
    pdfcreator={},   
    pdfproducer={}, 
    pdfkeywords={quantum scars}, 
    pdfnewwindow=true,      
    colorlinks=true,       
    linkcolor=black,          
    citecolor=blue,        
    filecolor=magenta,      
    urlcolor=blue           
}
\newcommand{\beq}{\begin{equation}}
\newcommand{\eeq}{\end{equation}}

\newcommand{\ket}[1]{\left| #1\right\rangle}

\begin{document}

\title{Quantum Many-Body Scars and Weak Breaking of Ergodicity}
\author{Maksym Serbyn$^1$, Dmitry A. Abanin$^2$, and Zlatko Papi\'c$^3$}
\affiliation{$^1$IST Austria, Am Campus 1, 3400 Klosterneuburg, Austria}
\affiliation{$^2$Department of Theoretical Physics, University of Geneva, 24 quai Ernest-Ansermet, 1211 Geneva, Switzerland}
\affiliation{$^3$School of Physics and Astronomy, University of Leeds, Leeds LS2 9JT, United Kingdom}
\date{\today}

\begin{abstract}
Recent discovery of persistent revivals in quantum simulators based on Rydberg atoms have pointed to the existence of a new type of dynamical behavior that challenged the conventional paradigms of integrability and thermalization. This novel collective effect has been named \emph{quantum many-body scars} by analogy with weak ergodicity breaking of a single particle inside a stadium billiard.  In this overview, we provide a pedagogical introduction to quantum many-body scars and highlight the newly emerged connections with the semiclassical quantization of many-body systems. We discuss the relation between scars and more general routes towards weak violations of ergodicity due to ``embedded" algebras and non-thermal eigenstates, and highlight possible applications of scars in quantum technology.   
\end{abstract}

\maketitle

\section{Introduction}
A fundamental challenge in physics is to predict how systems evolve in time when taken out of their equilibrium state. The behavior of familiar classical systems, e.g., the clock pendulum, is perfectly periodic and thus possible to predict at late times. This is an example of \emph{integrable} behavior. Integrable systems, however, are special; under perturbations, their periodic behavior gradually disappears and ultimately gives way to \emph{chaos}. The microscopic chaotic motion allows an ensemble of coupled systems to reach the state of thermal equilibrium, which can be effectively described using the textbook methods of statistical mechanics.

While many open questions remain in the field of classical chaotic systems, the non-equilibrium \emph{quantum} matter is another frontier of modern research.  The intense focus on understanding routes to thermal equilibrium in isolated quantum systems with many degrees of freedom  is partly due to the advent of synthetic systems based on  atoms, ions, or superconducting circuits as elementary building blocks~\cite{Bloch2012,quantumsimulationRMP14}. The relaxation times of these systems are long compared to, e.g., those of electrons in solids,  and experiments allow to monitor quantum dynamics at the single-atom level, thus opening a window to non-equilibrium quantum phenomena.  These capabilities have renewed the interest in foundational questions of quantum physics, such as the emergence of statistical-mechanics description in an isolated many-body system. Provided parts of the system are able to act as heat reservoirs for its other parts, an initial non-equilibrium state relaxes to thermal equilibrium, with a well-defined effective temperature. Such systems are called thermal or quantum-ergodic.

Surprisingly, not all quantum systems are ergodic. 
For example, finely tuned one-dimensional systems~\cite{Kinoshita06} may fail to thermalize due to their rich symmetry structure known as quantum integrability. However, when such systems are weakly perturbed, ergodicity is restored. On the other hand, disorder in an interacting system may induce many-body localization (MBL)~\cite{Huse-rev, AbaninRMP}, a close relative of the celebrated Anderson localization for non-interacting particles. Similar to integrable systems, MBL systems have a macroscopic number of conservation laws, which prevent thermalization. These conservation laws, however, are robust with respect to perturbations, and MBL is an example of a phase of matter which embodies strong ergodicity breaking.

The behavior of quantum-ergodic systems is governed by the so-called  Eigenstate Thermalization Hypothesis (ETH)~\cite{DeutschETH,SrednickiETH}, a powerful conjecture which explains the process of thermalization at the level of the system's energy eigenstates. The ETH states that individual eigenstates of quantum-ergodic systems act as thermal ensembles, thus the system's relaxation does not depend strongly on the initial conditions. However, in 2018, an experiment on a new family of Rydberg-atom quantum simulators  revealed an unforeseen dynamical behavior~\cite{Bernien2017}. The experiment observed significant qualitative differences in the dynamics, depending on the choice of the initial state: while certain initial states showed relaxation to thermal ensembles, as expected in an ergodic system, other states  exhibited periodic revivals. Such revivals were surprising given that the system did not have any conserved quantities other than total energy and it was free of disorder, thus ruling out integrability and localization as possible explanations. What made a particular initial state find its way back in a Hilbert space with dimension in excess of $4\times 10^{10}$, for the experimental system of $51$ Rydberg atoms? 

A flurry of subsequent theoretical works have addressed the puzzle posed by the experiment: the nature and origin of this new regime of ergodicity breaking, intermediate between thermalization and strong ergodicity breaking. The key to understanding this new behavior was the discovery of anomalous, non-thermal eigenstates in the highly excited energy spectrum of the Rydberg atom system~\cite{Turner2017}. These eigenstates provide an example of what we will refer to as \emph{weak} violation of ETH: despite being strongly non-thermal, the anomalous eigenstates comprise a vanishing fraction of the Hilbert space and they are immersed in a much larger sea of thermal eigenstates. 

One may be puzzled as to why non-thermal eigenstates in the highly-excited spectrum would bear any importance. After all, the energy level spacing between highly-excited eigenstates in a many-body system decreases exponentially with the number of atoms, and preparing the system in a particular eigenstate is challenging, if not impossible. Therefore, such eigenstates might be expected to remain ``invisible" to any experiment performed in the lab. Nevertheless, it was shown that these non-thermal eigenstates indeed underpin the real-time dynamics observed in experiment, inspiring new connections between quantum many-body dynamics and studies of classical chaos~\cite{wenwei18TDVPscar}. By analogy with chaotic stadium billiards, which also host non-thermal eigenstates whose ergodicity breaking has been visualized as ``scars" of classical periodic orbits~\cite{Heller84}, the non-thermal eigenstates in the Rydberg atom chains have been dubbed ``quantum many-body scars"~(QMBSs). 

\begin{figure}[htb]
\includegraphics[width=\columnwidth]{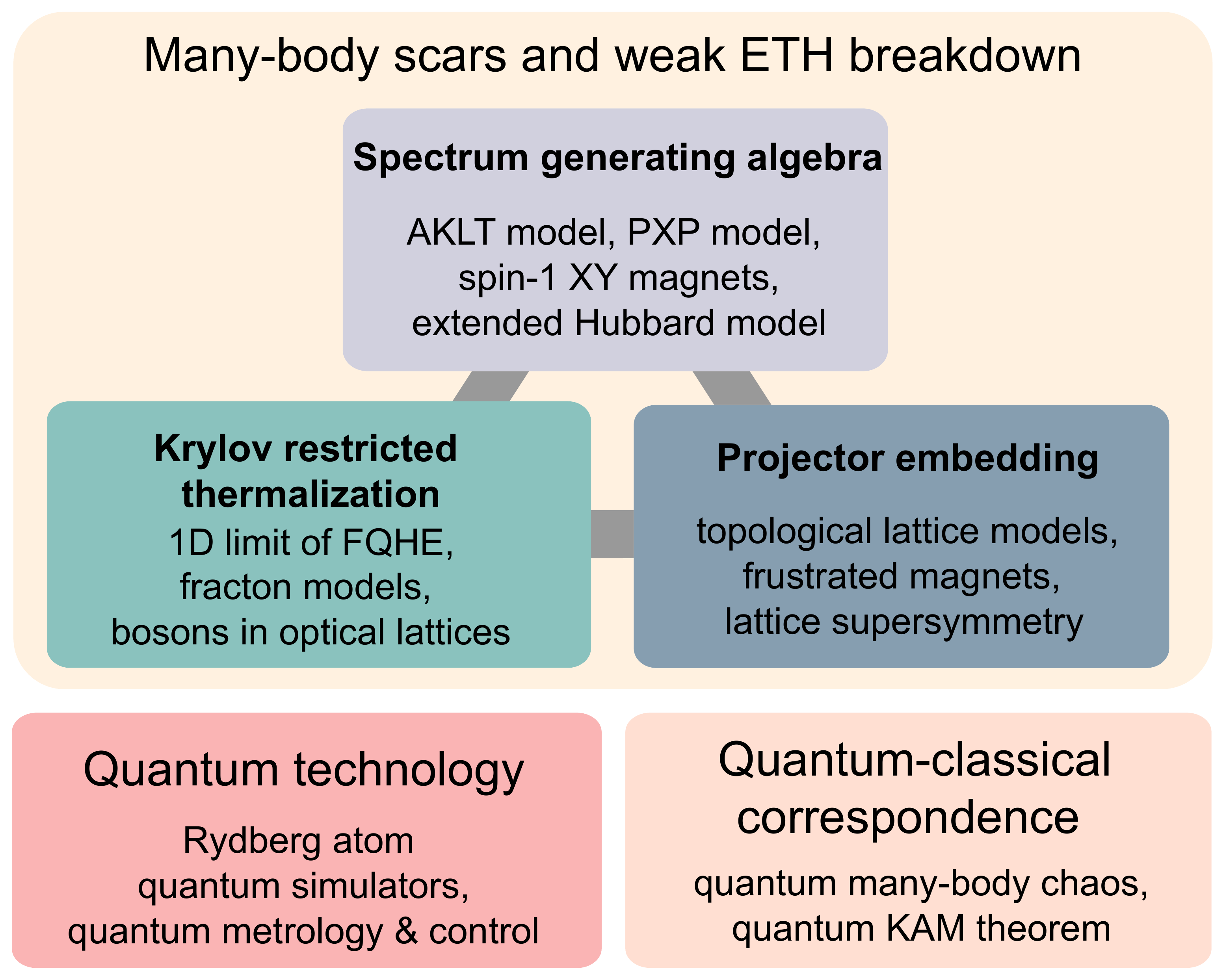}
\caption{ Top: Examples of physical systems where weak ergodicity breakdown  has recently been discovered. These systems can be classified according to three main mechanisms of many-body scarring, reviewed below. Bottom: Studies of many-body scarring have sparked interest in fundamental problems in the field of quantum many-body chaos and potential applications in quantum technology.
\vspace{20pt}
}
\label{Fig0} 
\end{figure}

The discovery of QMBSs has triggered  broader investigations of weak ergodicity breaking in a variety of quantum systems, as opposed to the strong ergodicity breaking by integrability and disorder. 
These studies have unearthed a rich landscape of many-body scarred models
with universal algebraic structures, shedding new light on the well-known condensed matter models such as the Hubbard model and the Affleck-Kennedy-Lieb-Tasaki (AKLT) model.  In this review, we survey the recent progress in weak ergodicity breaking phenomena by highlighting relations between different realizations of non-thermal dynamics, schematically illustrated in Fig.~\ref{Fig0}. As we explain in detail below, known scarred models are characterized by a subspace which  is decoupled from the rest of the energy spectrum and cannot be attributed to a symmetry of the system. The origin of this subspace can vary depending on the model, and a few common mechanisms, highlighted in Fig.~\ref{Fig0}, will be presented below. Furthermore, we discuss novel theoretical approaches that aim to bridge quantum and classical chaos in many-body systems, and the representation of scarred dynamics  in terms of tensor networks.  Finally, we outline promising directions for future research, putting an emphasis on emerging connections between weak ergodicity breaking and other fields. We mention other routes to weak ergodicity breaking, such as mesons in theories with confinement, the effects of additional conservation laws that fracture the Hilbert space, and frustration-driven glassy behavior, whose relation to QMBS yet remains to be fully understood. We conclude with a brief discussion of potential applications of QMBS in quantum sensing and metrology.

\section{Phenomenology of quantum scarring}

In experiment, quantum thermalization  is conveniently probed by quenching the system: one prepares a non-equilibrium initial state $|\psi(0)\rangle$, which is typically short-range correlated (e.g., a product state of particles), and monitors its fate after time $t$. For cold atoms and trapped ions, which are well-isolated from any thermal bath, the system can be assumed to evolve according to the Schr\"odinger equation for the system's Hamiltonian~$H$.  The process of thermalization is monitored by measuring the time evolution of local observables, $\langle O(t) \rangle$, as the dynamics explores progressively larger parts of the system's energy spectrum. The results of such measurements are remarkably different in  typical quantum-ergodic systems compared to non-ergodic localized systems. In ergodic systems, following a brief initial transient, local observables relax to their thermal value and stay near that value at later times. This behavior is reminiscent of classical chaotic systems, which effectively ``forget" their initial condition. In contrast, in MBL systems local observables reach a stationary value which is non-thermal, retaining the memory of the initial state -- a hallmark of broken ergodicity. 

It is important to emphasize that the dynamical behavior of ergodic or non-ergodic systems mentioned above has been verified to hold for quenches from \emph{any} physical initial states. This typicality was recently found to break down in  experiments on arrays of Rydberg atoms~\cite{Bernien2017}. The building block of these experiments is an individual Rydberg atom, which may be viewed as an effective two level system, where the two states  $|{\circ}\rangle$,  $|{\bullet}\rangle$ correspond, respectively, to an atom in the ground state and an atom in the so-called Rydberg state. 
When subject to a microwave field, each atom undergoes Rabi oscillations, $|{\circ}\rangle \leftrightarrow |{\bullet}\rangle$, freely flipping between its two states. However, when assembled in an array,  the  atoms in $\ket{{\bullet}}$ states interact via repulsive van der Waals force, whose strength is strongly dependent on the distance between the atoms. By tuning the inter-atom distance one can achieve the regime of the Rydberg blockade~\cite{Labuhn2016} where the excitations of neighboring atoms, e.g., $|\cdots{\bullet}{\bullet}\cdots\rangle$, are energetically prohibited. This makes the system kinetically constrained as each atom is only allowed to flip if all of its neighbors are in $|\circ\rangle$ state.

\begin{figure}[tb]
\includegraphics[width=\columnwidth]{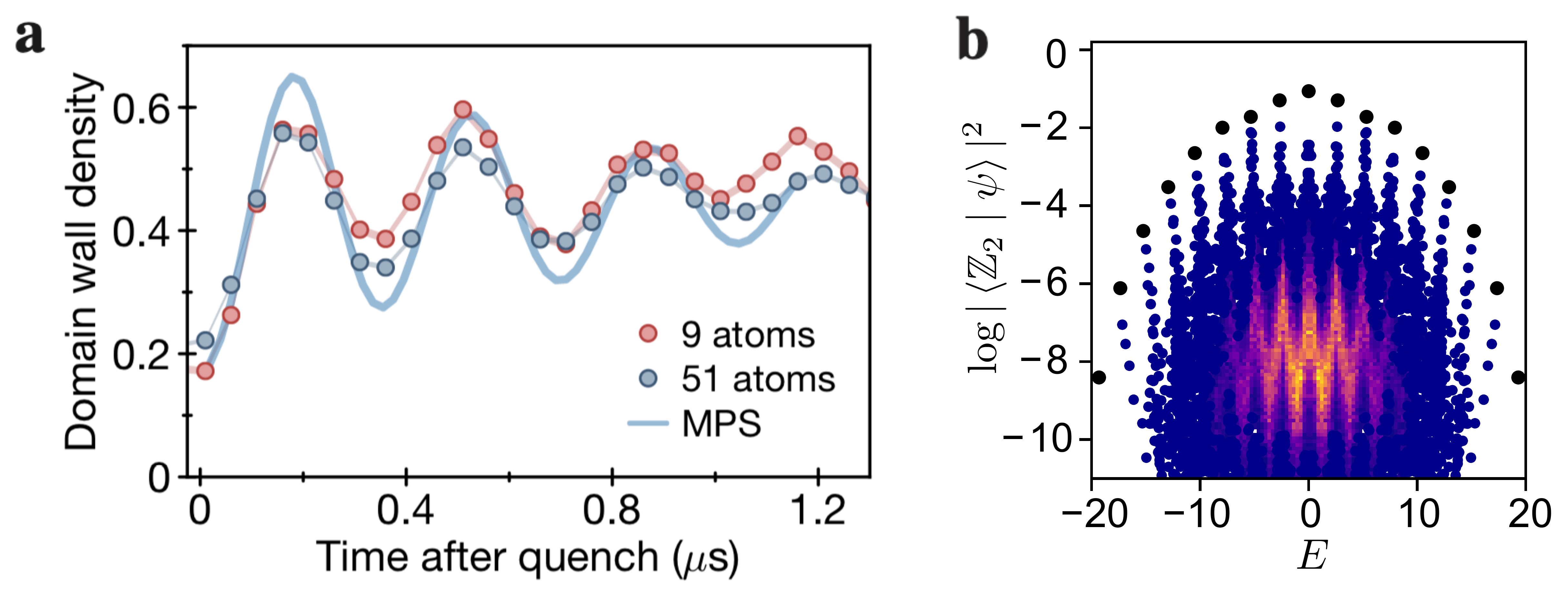}
\caption{{\bf Weak breakdown of thermalization.} 
{\bf a:} Observation of persistent revivals in 
the domain-wall density when Rydberg-atom quantum simulators are quenched from $\ket{\mathbb{Z}_2}$ initial state~\cite{Bernien2017}. 
The oscillations were found to persist in large chains up to $51$ atoms, and they were reproduced in the dynamical simulations using matrix product state (MPS) methods. 
{\bf b:} Weak breakdown of thermalization in Rydberg atom arrays was theoretically attributed to atypical eigenstates -- quantum many-body scars -- which are distinguished by their anomalously enhanced overlaps with $\ket{\mathbb{Z}_2}$ state~\cite{Turner2017}. Each dot represents an eigenstate of energy $E$ in the PXP model in Eq.~(\ref{Eq:PXP}) with  $L=32$ atoms. Color scheme illustrates the density of points.
\vspace{20pt}
}
\label{Fig1} 
\end{figure}

The dynamics of large one-dimensional arrays of Rydberg atoms has been probed by studying quenches from a period-2 density wave initial state, $|\mathbb{Z}_2 \rangle \equiv |{\bullet}{\circ}{\bullet}{\circ}\ldots\rangle$, which is the state containing the maximal possible number of Rydberg excitations allowed by the blockade.  The dynamics of domain wall density, where domain walls are defined as adjacent $|{\circ}{\circ}\rangle$ or $|{\bullet}{\bullet}\rangle$ configurations (the latter is excluded in the regime of perfect blockade), revealed long-time oscillations shown in Fig.~\ref{Fig1}{\bf a}. Such robust oscillations were surprising, given that the state $|\mathbb{Z}_2\rangle$ is not a low-energy state, but rather corresponds to an infinite-temperature ensemble for the atoms in the Rydberg blockade. Thus the quench dynamics is not \emph{a priori} limited to a small part of the energy spectrum, and cannot be attributed to weakly interacting low-energy excitations in the system. Moreover, the oscillations could not be explained by any known conservation law in the system, and their frequency did not coincide with the bare Rabi frequency, signalling the importance of many-body effects. 

The experimental findings have been corroborated by theoretical studies  of the idealized model of atoms in the Rydberg blockade -- the so-called  ``PXP" model~\cite{Lesanovsky2012}, 
\begin{equation}\label{Eq:PXP}
H_{\mathrm{PXP}} = \sum_i P_{i-1}\sigma^x_i P_{i+1}.
\end{equation}
Here $\sigma_i^x = {|\circ\rangle}_i { \langle \bullet |}_i +  {|\bullet  \rangle}_i {\langle\circ |}_i $ denotes the Pauli $x$ matrix responsible for Rabi oscillations of an individual atom on site $i$, and projectors onto the ground state, $P_i  = |{\circ}\rangle_i\langle{\circ}|_i$, enforce the Rydberg blockade and make the model interacting.  Numerical studies have observed that the PXP energy levels repel, thus confirming that the model is non-integrable~\cite{Turner2017}. Further, simulations of the quench dynamics have demonstrated revivals of the wave function and local observables when the system is quenched from $|\mathbb{Z}_2\rangle$ state~\cite{Sun2008, LesanovskyDynamics} as well as the period-3 density wave state, $|\mathbb{Z}_3 \rangle \equiv |{\bullet}{\circ}{\circ}{\bullet}{\circ}{\circ}\ldots\rangle$. In contrast, other initial states, such as  $|{0}\rangle = \ket{{\circ}{\circ}{\circ}\ldots}$ state and period-4 density wave state, showed fast relaxation without revivals~\cite{TurnerPRB}. Similar  phenomenology has also been identified in the two-dimensional PXP model~\cite{Michailidis2D,Hsieh2020}, and in related models of transverse Ising ladders~\cite{Voorden2020} and the periodically driven PXP model~\cite{Sugiura2019,Mizuta2020, Mukherjee2020}. This special behavior of the PXP model was found to be robust to certain perturbations~\cite{Lin2020} including disorder~\cite{MondragonShem2020}, while other perturbations destroy it by making the PXP model integrable~\cite{FendleySachdev}, frustration-free~\cite{LesanovskyMPS} or thermalizing~\cite{TurnerPRB}.

An important  step in understanding the origin of the  observed oscillations was made  by theoretical studies of many-body eigenstates of the PXP model. Figure~\ref{Fig1}{\bf b} reveals a number of ``atypical" eigenstates in the spectrum of the PXP model, which are distinguished by their high overlap with the same $\ket{\mathbb{Z}_2}$ state used for initiating the quench in Figure~\ref{Fig1}{\bf a}. For a quantum-ergodic system, the ETH stipulates that these overlaps should be a smooth function of energy; by contrast, the strong enhancement of overlaps of \emph{certain} eigenstates compared to others at the same energy $E$, seen in Figure~\ref{Fig1}{\bf b}, signals a breakdown of the ETH. This is a weak violation of the ETH since the number of special eigenstates was found to scale linearly with the number of atoms, in contrast to an exponentially larger number of thermalizing eigenstates. Moreover, the special eigenstates are ``embedded"  at roughly equidistant energies
in an otherwise thermalizing spectrum of the PXP model, thus accounting for the  dynamical revivals in Figure~\ref{Fig1}{\bf a} with a single dominant frequency. We note that various schemes for constructing some of these atypical eigenstates in the PXP model, both exact~\cite{lin2018exact} and approximate~\cite{TurnerPRB,Iadecola2019}, have been devised. 

The core phenomenology of the PXP model -- the \emph{small} number of ETH-violating eigenstates within the thermalizing spectrum and the presence of many-body revivals and slow relaxation in quenches from \emph{specific} initial states -- bears parallels with the physics of a single particle confined to a stadium-shaped billiard. In the latter case, the dynamics of a quantum wave packet is known to be sensitive to the initial conditions, which was attributed to the existence of unstable periodic orbits when the billiard is made classical by sending $\hbar\to 0$ (see Box~1). Initialization of a quantum wave packet on or near such an orbit leads to dynamical recurrences where the particle tends to cluster around the orbit, a phenomenon called \emph{quantum scarring}~\cite{Heller84}. Moreover, the periodic orbit leaves an imprint on the eigenfunctions of a particle, which exhibit anomalous concentration in the vicinity of the periodic orbit, rather than being spread uniformly across the billiard.

While the analogy between revivals in the PXP model  and scars in billiards is indeed suggestive, making it more precise has required new theoretical insights. As we discuss below, these approaches have followed two complementary strands. The first approach, discussed in Section~\ref{Sec:ETH}, focuses on constructing non-thermal eigenstates in many-body quantum systems. This approach has been particularly fruitful in identifying a large family of scarred many-body systems, some of which were highlighted in Fig.~\ref{Fig0}.  An alternative perspective, which focuses on parallels between quantum many-body dynamics and  classical dynamical systems,  will be discussed in Section~\ref{Sec:Chaos}.

\onecolumngrid
\definecolor{shadecolor}{rgb}{0.8,0.8,0.8}
\begin{shaded}
\noindent{\bf Box 1 $|$ Scars in quantum billiards}
\end{shaded}
\vspace{-9mm}
\definecolor{shadecolor}{rgb}{0.9,0.9,0.9}
\begin{shaded}
\noindent What can we learn about the behavior of a  quantum system from looking at its classical counterpart? The  Bohr-Sommerfeld quantization demonstrates that classical integrable systems, such as harmonic oscillator or hydrogen atom, have very special quantum spectrum and eigenstates.  For classically chaotic systems, such as the Bunimovich stadium in Fig.~B1 below, this  quantization method does not work. Typically, quantum counterparts of classically chaotic systems display level repulsion in their spectrum and their eigenstates look random.  However, short unstable periodic orbits may leave a strong imprint on the system's quantum dynamics and eigenstate properties  -- this influence of classical periodic orbits is called ``quantum scarring"~\cite{HellerLesHouches}: \emph{A quantum eigenstate of a classically chaotic system has a  `scar' of a periodic orbit if its density on the classical invariant manifolds near the periodic orbit  is enhanced over the statistically expected density.} Scarring is a weak form of ergodicity breaking since the individual eigenstates of the billiard are ``almost always" ergodic as the phase space area affected by scarring vanishes in the semiclassical limit $\hbar\to 0$, shrinking around the periodic orbit.

How does one detect scars? Most simply, by visualizing the wave function probability density, as in the case of Bunimovich stadium in Fig.~B1{\bf c}-{\bf d}. The probability density can be directly  compared with the corresponding classical orbits in Fig.~B1{\bf a}-{\bf b}. Moreover, scars leave an imprint on the dynamics: a wave packet launched in the vicinity of an unstable periodic orbit will tend to cluster around the orbit at later times, displaying larger return probability than a wave packet launched elsewhere in phase space. This is seen in the autocorrelation function in Fig.~B1{\bf e}. Furthermore, such a wave packet can be expanded over a small number of eigenstates that have approximately similar energy spacing -- see Fig.~B1{\bf f} -- in contrast to an arbitrary wave packet.   

What is the significance of quantum scars?
 First, scars provide a counterexample to the intuitive expectation that every eigenstate of a classically chaotic system should locally look like random superposition of plane waves~\cite{Berry1983}. Second, also counterintuitively, a scarred quantum system appears to bear a stronger influence of the periodic orbit compared to its classical counterpart, since in the latter case there is no enhancement of density along the periodic orbit in the long-time limit.  Finally, scars play a role in many physical systems, including microwave cavities~\cite{Sridhar1991}, semiconductor quantum wells~\cite{Wilkinson1996}, and the hydrogen atom in a magnetic field~\cite{Wintgen1989}.

\begin{center}
\label{fig:billiard}
\includegraphics[width=0.9\columnwidth]{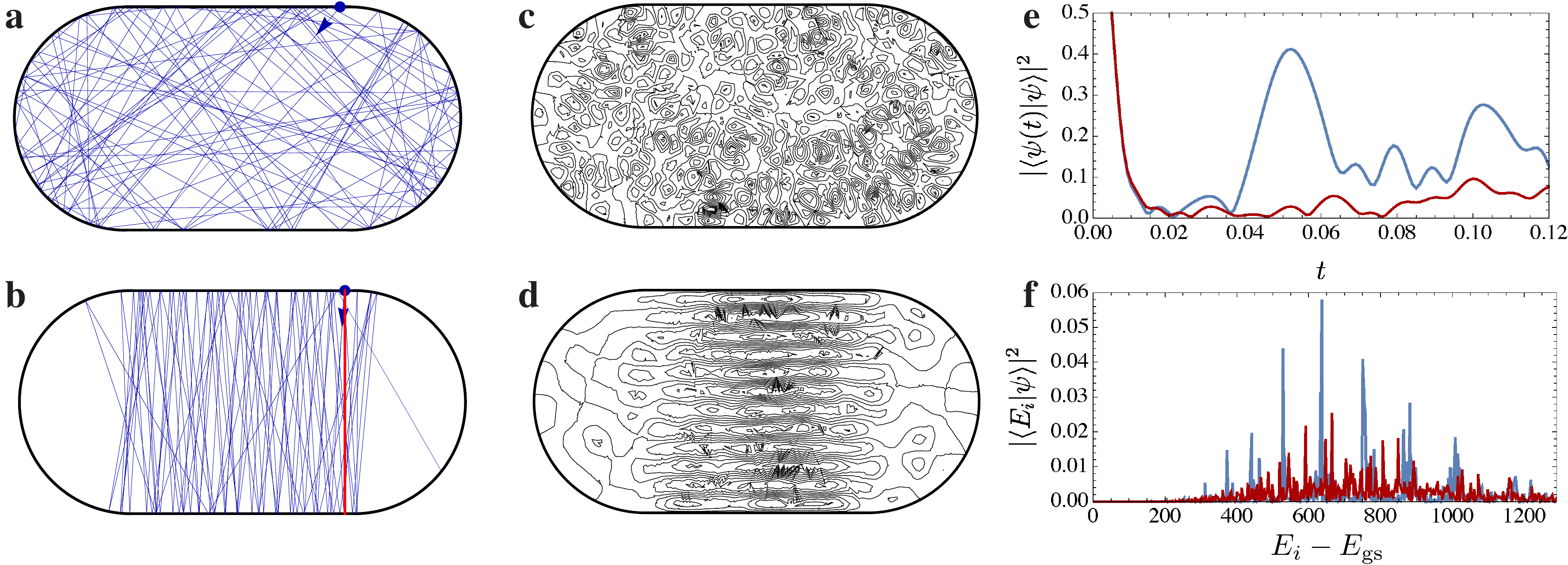}
\end{center}
\small {\noindent{\bf Figure B1 $|$ Scars in a stadium billiard.} 
{\bf a:} Classical particle initialized away from an unstable periodic trajectory displays chaotic motion.  {\bf b:} In contrast, when launched near an unstable periodic trajectory shown in red, the particle spends a long time in its vicinity before escaping. {\bf c:} The probability density of a typical highly excited eigenstate of the billiard resembles a collection of random plane waves. {\bf d:} Probability density of a quantum-scarred eigenstate looks very different from a collection of plane waves, instead being strongly concentrated near the periodic trajectory. The eigenstates of the billiard are obtained by solving the Schr\"odinger equation with the wave function vanishing at the boundary. {\bf e-f:} Autocorrelation function of a Gaussian wave packet launched vertically at the center of billiard shows pronounced revivals (blue line) and the expansion of the initial state over the billiard eigenstates  reveals a periodic sequence of peaks. In contrast, the wave packet launched at $45^\circ$ angle does not revive and features a continuum of frequencies (red lines).} 
\end{shaded}

\twocolumngrid

\section{Mechanisms of weak ergodicity breaking \label{Sec:ETH}}

 \begin{figure*}[t]
\includegraphics[width=2\columnwidth]{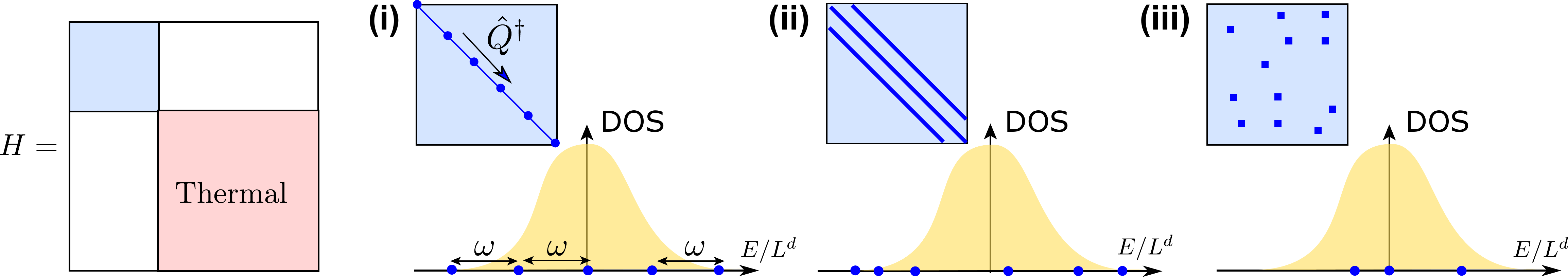}
\caption{{\bf Embeddings of scarred eigenstates in a many-body system.}  The Hamiltonian matrix $H$, acting on the full Hilbert space, is schematically split into the non-ergodic subspace (blue box) and the remaining thermal subspace (red box). Panels {\bf (i)}-{\bf(iii)} illustrate possible mechanisms underlying the emergence of the decoupled subspace. Density of states (DOS) schematically shows the positions of scarred eigenstates (blue dots) within the continuum of thermalizing states. {\bf (i)} Exact scars due to a spectrum generating algebra and resulting eigenstates which are equidistant in energy. {\bf (ii)} Exact Krylov subspace represented by a tridiagonal matrix. {\bf (iii)} General subspace resulting from projector embedding. In any of these cases, the scarred subspace need not be \emph{exactly} decoupled from the thermalizing bulk, like in the PXP model. 
 } 
 \label{fig:mechanisms}
 \end{figure*}

\begin{figure*}
\includegraphics[width=1.9\columnwidth]{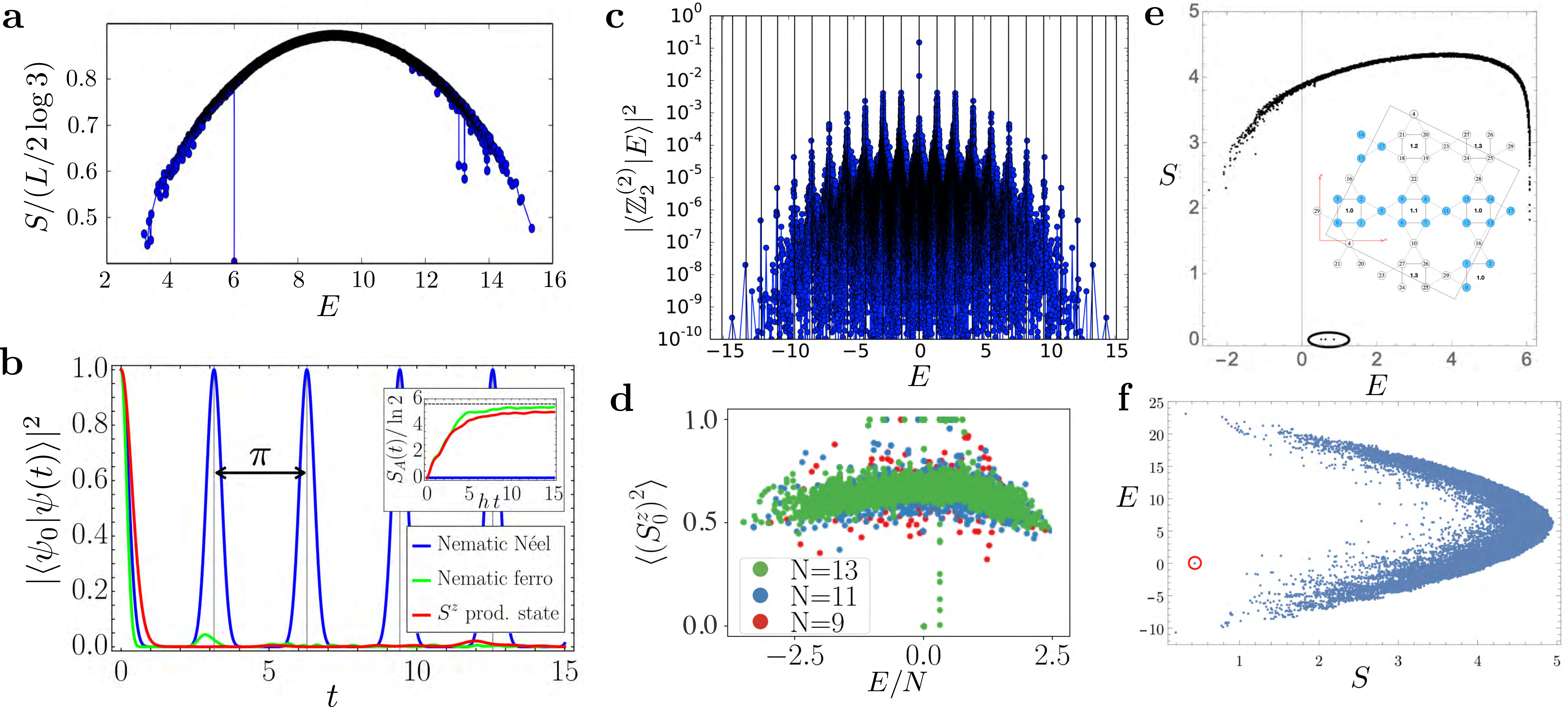}
\caption{
{\bf Mechanisms and physical realizations of weak ergodicity breaking.} {\bf a}-{\bf b}: Many-body scars due to spectrum generating algebras. {\bf a}: Exact scars in the AKLT model distinguished by their low entanglement entropy $S$~\cite{BernevigEnt}. {\bf b}: Perfect fidelity revival due to exact scars in a spin-1 XY magnet~\cite{Iadecola2019_2}. 
{\bf c}-{\bf d}: Krylov subspaces in fracton-like models. The model of the fractional quantum Hall effect on a thin cylinder~\cite{Moudgalya2019} supports scarred eigenstates and revivals from a product state ({\bf c}), while local observables in a spin-1 model with dipole conservation symmetry~\cite{Sala2019} weakly violate the ETH ({\bf d}).
{\bf e}-{\bf f}: Projector embedding of scarred states into thermalizing spectra. {\bf e}: Geometrical frustration on a 2D square-kagome lattice (inset) allows to embed weakly-entangled states of localized magnons (circled)~\cite{McClarty2020}. {\bf f}: Deformations of topological models, such as the 1D cluster model,  allows to embed scarred states (circled)~\cite{NeupertScars}. The same method applies to 2D toric code and 3D X-cube models. 
 }
\label{fig5} 
\end{figure*}

In recent works, non-thermal eigenstates have been theoretically identified in a number of non-integrable quantum models, revealing a plethora of scarred many-body systems.  A unifying property of all these systems  is the emergence of  a decoupled subspace within the many-body Hilbert space, not related to any global symmetry of the full Hamiltonian, spanned by the non-thermal eigenstates. Formally, the Hamiltonian of such systems is effectively decomposed as 
\begin{equation}\label{Eq:decompose}
H \approx H_{\mathrm{scar}} \bigoplus H_{\mathrm{thermal}},
\end{equation}
 where $H_\mathrm{scar}$ is the scarred subspace which is (approximately, or even exactly) decoupled from the thermalizing subspace, $H_{\mathrm{thermal}}$. The eigenstates that inhabit the subspace $H_\mathrm{scar}$ violate the ETH and have different properties compared to the majority of thermal eigenstates in $H_{\mathrm{thermal}}$.  Below we elucidate and contrast three  possible mechanisms (i)-(iii) that lead to such a decoupled subspace, whose summary is given in Fig~\ref{fig:mechanisms}. As we explain next, these mechanisms are realized by diverse physical systems illustrated in Fig.~\ref{fig5}.

{\bf (i)  Spectrum-generating algebra.} A family of highly-excited non-thermal eigenstates can be constructed systematically using the ``spectrum generating algebras", originally introduced in the context of high energy physics~\cite{Arno1988} and subsequently applied to the Hubbard model~\cite{YangEta,ZhangEta}. The first exact construction of many-body scarred eigenstates using this method has been achieved in the Affleck-Kennedy-Lieb-Tasaki (AKLT) model~\cite{BernevigEnt}, extending earlier work of Ref.~\cite{Arovas1989}.  In the AKLT model, as well as in several other non-integrable spin chain models~\cite{Iadecola2019_2, Iadecola2019_3,Chattopadhyay,OnsagerScars}, the spectrum generating algebra allows to \emph{exactly} construct scarred eigenstates, which can physically be interpreted as quasiparticle condensates built on top of the system's ground state.

The spectrum generating algebra is defined by  a local operator $\hat Q^\dagger$ which obeys the relation $([H, \hat Q^\dagger] - \omega \hat Q^\dagger) W = 0$, where $\omega$ is some energy scale and $W$ is a linear subspace of the full Hilbert space left invariant under the action of~$Q^\dagger$. In the Hubbard model, the subspace $W$ coincides with the full Hilbert space and $Q^\dagger$ assumes the role of an $\mathrm{SU}(2)$ ``eta-pairing" symmetry~\cite{Vafek}. 
The scenario relevant for many-body scarring is when $W$ is not the full Hilbert space and  $Q^\dagger$ is not associated with a symmetry of the Hamiltonian~\cite{MotrunichTowers},  e.g., as realized in the extended Hubbard model~\cite{MarkHubbard,MoudgalyaHubbard}. 
 Then, starting from some eigenstate $|\psi_0\rangle$ of $H$, the tower of states, $(\hat Q^\dagger)^n \vert \psi_0 \rangle$, are all eigenstates of $H$ with energies $E_0+n\omega$.  The eigenstates $(\hat Q^\dagger)^n \vert \psi_0 \rangle$ are candidates for many-body scars because their properties can be strongly non-thermal, provided $|\psi_0\rangle$ is sufficiently ``simple" (e.g., the ground state of $H$ if the latter is gapped) and $\hat Q^\dagger$ is a local operator (e.g., a sum of on-site spin flips in quantum spin chain models).

Often, $\hat Q^\dagger$ is interpreted as creating a wave packet corresponding to a ``quasiparticle" excitation (e.g., a magnon), and repeated applications of $\hat Q^\dagger$ create a condensate of such quasiparticles.  In a class of frustration-free models which include  the AKLT~\cite{MoudgalyaFendley}, the quasiparticle condensates are non-thermal and violate the ETH. A convenient diagnostic for the ETH violation and identifying scarred eigenstates is the entanglement entropy $S$, i.e., the von Neumann entropy of the reduced density matrix describing a subsystem of the full system in a given eigenstate. In ETH systems, the entanglement entropy of eigenstates scales extensively with the size of the subsystem. By contrast, scarred eigenstates have much lower entanglement, thus they can be detected as entropy-outliers, as seen in Fig.~\ref{fig5}{\bf a} for the AKLT model. Another important clue when looking for scarred eigenstates is that they appear at energies which are rational multiples of $\omega$.

In related approaches, scarred subspace may be conveniently engineered if $H$  has a certain symmetry, such as in a spin-1 XY magnet~\cite{Iadecola2019_2}. This model also hosts perfect wave function revivals from certain easily preparable initial states, as illustrated by its quantum fidelity, $|\langle \psi_0|\psi(t)\rangle|^2$, in Fig.~\ref{fig5}{\bf b}. Further variations of the algebraic construction have been given for spin-1 magnets~\cite{Chattopadhyay}, a spin-$1/2$ model with emergent kinetic constraints~\cite{Iadecola2019_3}, a spin chain with the dynamical Onsager algebra~\cite{OnsagerScars}, and a two-dimensional  frustrated  spin-1/2  kagome antiferromagnet~\cite{Lee2020}. Recent works have provided a more general framework for the construction of  non-thermalizing subspaces using group theory~\cite{Pakrouski2020, Ren2020,Dea2020}. It has also been pointed out that spectrum generating algebra can arise in open quantum systems, e.g., in the presence of dissipation or periodic driving~\cite{Buca2019,BucaHubbard}.

Finally, the PXP model in Eq.~(\ref{Eq:PXP}) also realizes an embedded algebra, in this case an approximate su(2) spin algebra~\cite{Choi2018}. The spin raising operator $H^+$ creates a Rydberg excitation anywhere on the even sublattice and removes an excitation anywhere on the odd sublattice, while respecting the Rydberg constraint. Similarly, the spin lowering operator $H^-$ performs the same process with the sublattices exchanged. The reason for this choice of $H^\pm$ is that their commutator defines the $z$-projection of spin,  $H^z \equiv \frac{1}{2} [H^+, H^-]$, for which $|\mathbb{Z}_2\rangle$ plays the role of the extremal weight state.  In principle, using this construction, one could form the Casimir operator for the su(2) algebra and obtain the tower of non-thermal eigenstates by acting repeatedly with  $H^+$ on the ground state of the Casimir, mirroring the general procedure outlined previously. However, this procedure is not analytically tractable due to the fact that the ground state of the Casimir operator is not known and the algebra of the PXP model is only approximate (recent works, however, have shown that weak deformations of the PXP model make the algebra and dynamical revivals progressively more accurate~\cite{Choi2018, Khemani2018,Bull2020}). Instead,  different schemes have been used to approximate scarred eigenstates in the PXP model starting from $|\mathbb{Z}_2\rangle$ product state~\cite{TurnerPRB,Turner2020}. 

{\bf (ii) Krylov restricted thermalization.} A spectrum generating algebra relies on the tower operator $Q^\dagger$ to construct the non-thermalizing subspace. In practice, finding such operators is likely to be limited to cases where they can be expressed in a simple local form. We next describe a related mechanism for producing exactly embedded subspaces, which does not require \emph{a priori} knowledge of $Q^\dagger$. 

For a Hamiltonian $H$ and some arbitrary vector in the Hilbert space, $|\psi_0\rangle$, the \emph{Krylov subspace}, $\mathcal{K}$, is defined as a set of all vectors obtained by repeated action of $H$ on $|\psi_0\rangle$, i.e., $\mathcal{K}=\mathrm{span}\{ |\psi_0\rangle, H|\psi_0\rangle, H^2 |\psi_0\rangle, \ldots\}$.  
In numerical linear algebra,  subspace $\mathcal{K}$ is routinely used in iterative methods for finding extremal eigenvalues of a large matrix $H$, such as the Arnoldi and Lanczos algorithms. A special case of particular interest is when $\mathcal{K}$ happens to be \emph{finite}, i.e., the sequence of Krylov vectors terminates after $n+1$ steps: $H^{n+1}|\psi_0\rangle=0$. This causes a dynamical 	``fracture" of the Hilbert space in the sense that the Schr\"odinger dynamics initialized in any state $|\psi\rangle \in \mathcal{K}$ must remain within the same subspace at any later time,  $\exp(-\frac{i}{\hbar}tH)|\psi\rangle \in \mathcal{K}$, hence the name ``Krylov restricted thermalization"~\cite{MoudgalyaKrylov}.  Note that $\mathcal{K}$ can be as small as one-dimensional or exponentially large in system size, and it can be either integrable or thermalizing. By performing a Gram-Schmidt orthogonalization, we can transform $\mathcal{K}$ into a tridiagonal matrix, which will be perfectly decoupled from the rest of the spectrum of $H$. 

We distinguish the case of the Krylov subspace from the previous case of the algebra subspace because, in general, it can be difficult to analytically diagonalize the tridiagonal matrix and construct the corresponding  $\hat Q^\dagger$. Note that a tridiagonal Krylov subspace does not guarantee revivals, even if the root state $|\psi_0\rangle$ is experimentally preparable. This is because general tridiagonal matrices do not support revivals, unless their matrix elements are tuned to special values. Moreover, depending on the model, the Krylov subspace may only be \emph{approximately} decoupled, i.e., there may exist small matrix elements that connect it with the thermalizing bulk. 

Realizations of Krylov subspaces occur in models of the fractional quantum Hall effect in a quasi one-dimensional limit~\cite{Moudgalya2019} and in models of bosons with constrained hopping on optical lattices~\cite{bosonScars, Zhao2020}. In the former case, shown in Fig.~\ref{fig5}{\bf c}, $\mathcal{K}$ is thermalizing but also supports scarred eigenstates and revivals from a particular density-wave product state. Scarred towers in overlaps with a density-wave state in Fig.~\ref{fig5}{\bf c} signal a weak ETH violation due to anomalous concentration of eigenstates in the Hilbert space, analogous to the case of PXP model in Fig.~\ref{Fig1}{\bf b}.  

More generally, the Krylov fracture was shown to be present in both Hamiltonian models~\cite{Sala2019} and random-circuits~\cite{Pretko2019,Khemani2019_2} which feature a conservation of dipole moment -- a symmetry characteristic of the so-called fracton topological phases. An example of weak breaking of the ETH in a one-dimensional fracton-like model~\cite{Sala2019} is shown in Fig.~\ref{fig5}{\bf d}, which shows the matrix elements of a local spin observable in the system's eigenstates. Once again, the large spread of these matrix elements and their nonmonotonic behavior as a function of energy indicate a weak violation of the ETH.

{\bf (iii) Projector embedding.} At the highest level of abstraction, we can ask if one could embed an arbitrary subspace into the spectrum of a thermalizing system? For concreteness, assume we are given an arbitrary set of states $ \vert \psi_i \rangle$ that span our target scar subspace $H_{\mathrm{scar}}=\mathrm{span}\{  \vert \psi_i \rangle\}$, which we wish to embed into a thermalizing Hamiltonian $H$ as in Eq.~(\ref{Eq:decompose}). This can be achieved via the ``projector embedding" construction due to Shiraishi and Mori~\cite{ShiraishiMori}, which has proven to be a versatile method for constructing scarred states in diverse models ranging from lattice supersymmetry~\cite{Surace2020} to flat bands~\cite{Kuno2020}.

Projector embedding assumes that our target states $\vert \psi_i \rangle$, being non-thermal, are annihilated by local projectors, $P_i \vert \psi_j \rangle = 0$, for any $i$ ranging over lattice sites $1,2,\ldots,L$. 
Next, consider a lattice Hamiltonian of the form
$H = \sum_{i=1}^L P_i h_i P_i + H^{\prime}$,
where  $h_i$ are arbitrary operators that are have support on a finite number of sites around $i$, and $[H^{\prime},P_i] = 0$ for all  $i$. It follows that
$P_i H \vert \psi_j \rangle = P_i H^{\prime} \vert \psi_j \rangle = H^{\prime} P_i \vert \psi_j \rangle = 0$,
thus $[H,\sum_i P_i]=0$. Therefore, $H$ takes the desired block diagonal form in Eq.~(\ref{Eq:decompose}). The target eigenstates can in principle be embedded at arbitrarily high energies for a suitably chosen $H'$, which also ensures that the model is overall non-integrable~\cite{ShiraishiMori}. However, there is no guarantee that the embedded states must be equidistant in energy and they may even be degenerate, such that this scheme could produce models which do not exhibit wavefunction revivals. 

Physical mechanisms such as geometrical frustration in correlated materials can  implement the projector embedding. Fig.~\ref{fig5}{\bf e} shows an example of a 2D square-kagome lattice model, where scarred states representing localized magnons can be embedded into the highly-excited energy spectrum~\cite{McClarty2020}. Moreover, projector embedding naturally lends itself to physical realization in various topologically-ordered lattice models~\cite{Wildeboer2020}, which are themselves defined in terms of projectors. Deformations of such models~\cite{NeupertScars}, which include the 2D toric code model and 3D X-cube model, can be used to embed scarred states, as shown in Fig.~\ref{fig5}{\bf f}.

In summary, we have presented three mechanisms (i)-(iii) that can give rise to non-thermal eigenstates in the spectrum of an otherwise thermalizing model.  We emphasize that these mechanisms  are not mutually exclusive,  they rather underscore how weak ergodicity breaking is manifested in various models. The connections between different mechanisms are most saliently illustrated by the PXP model. The PXP model, with its generalizations to higher spins~\cite{wenwei18TDVPscar} and quantum clock models~\cite{Bull2019},  falls into category (i), yet its algebra is only \emph{approximate}, and the subspace has weak couplings to the thermal bulk. In addition, the PXP model also realizes an \emph{approximate} Krylov subspace built upon $|\mathbb{Z}_2\rangle$ state (the subspace becomes exact when generated by $H^+$). Finally, a single PXP eigenstate at zero energy in the middle of the spectrum can be viewed as projector-embedded AKLT ground state~\cite{Shiraishi_2019}. While similar connections between scarring mechanisms may be anticipated in other models mentioned above, their demonstration remains an open problem. 

\section{Scars and periodic orbits in many-body systems \label{Sec:Chaos}}

Thus far we have surveyed a growing landscape of models which host ETH-violating scarred eigenstates. While the term ``quantum many-body scar'' is often used to denote general non-thermal eigenstates in a non-integrable system, a complete analogy with scars in quantum billiards necessitates some notion of a classical trajectory underlying the atypical eigenstates. This directly relates to the important problem in the field of quantum chaos: finding the classical counterpart of a many-body quantum system. Two commonly used methods -- mean-field theory and large-$N$ limit -- are not expected to work for the majority of quantum models discussed in Section~\ref{Sec:ETH}, as these models have small on-site Hilbert space, thus are far away from the mean-field limit. 
Progress on this question has recently been achieved using a variational framework based on matrix product states (MPS). Below we briefly outline this method and discuss the new physical insights it allowed to obtain about the PXP model -- the only available example at this stage. Afterwards we discuss the prospects of finding periodic trajectories in other scarred models and broader applications of the method beyond many-body scars.

Optimal projection of quantum dynamics onto a given variational manifold can be achieved by the time-dependent variational principle (TDVP) pioneered by Dirac~\cite{Dirac1930}. The variational principle determines the best direction of evolution within the manifold, such that the difference between exact and projected dynamics is minimal -- see Box~2 for details. It has been realized~\cite{Haegeman} that TDVP can be turned into a powerful computational tool for many-body systems when the manifold is parametrized by a family of MPS states. The advantage of MPS parametrization is that it naturally extends mean field-like product states by incorporating non-trivial but finite range correlations. In particular, the results of unitary time evolution from the initial product state can be efficiently represented as an MPS state. This MPS framework has recently been extended into a path integral over ``entanglement''-based degrees of freedom~\cite{Green2016},  and it has been used to calculate the diffusion coefficient in quantum many-body systems~\cite{Leviatan2017}.

\begin{figure}[t]
\includegraphics[width=0.9\columnwidth]{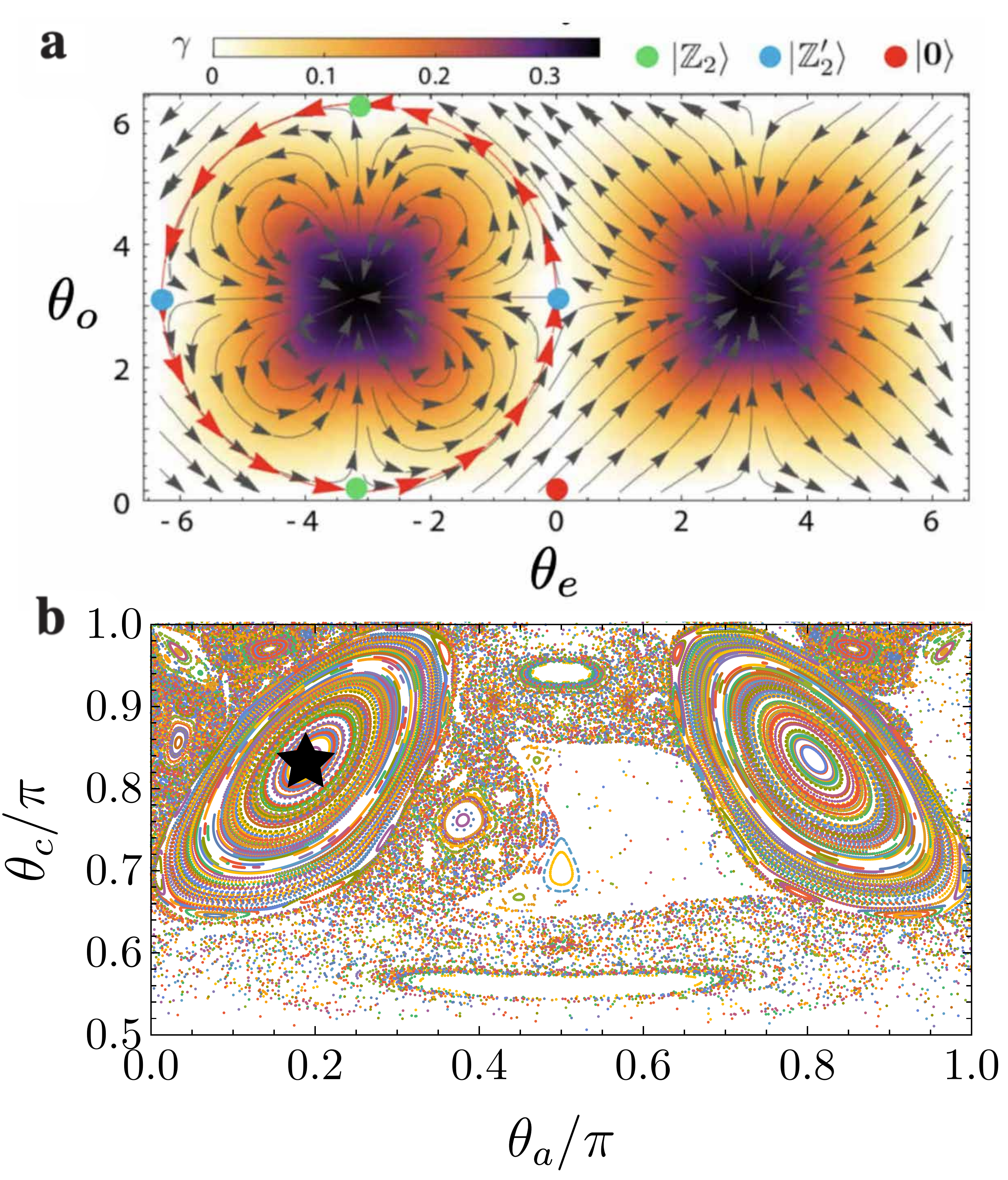}
\caption{{\bf Many-body scarring and periodic orbits in the semiclassical dynamics.} 
{\bf a:} Phase diagram obtained from the variational ansatz with two degrees of freedom reveals an unstable periodic trajectory, which is responsible for the revivals in the PXP model from $|\mathbb{Z}_2\rangle$  state~\cite{wenwei18TDVPscar}. Color scale shows quantum leakage which quantifies the instantaneous discrepancy between quantum and variational dynamics.
{\bf b:} Dynamical system with three angles $\theta_{a,b,c}$ results in a Poincar\'e section with many stable periodic trajectories (one of them denoted by the star) in centers of regular regions surrounded by the chaotic sea~\cite{Michailidis2020}.  This classical dynamical system describes, e.g., quantum revivals from $|\mathbb{Z}_3\rangle$ initial state in the PXP model.
}
\label{Fig:Ry} 
\end{figure}

The ability of MPS states to incorporate entanglement made it an ideal tool to explore semiclassical dynamics of the PXP model, which involves non-local correlations due to the Rydberg blockade. The variational MPS approach is tailored to capture dynamics from the initial $|\mathbb{Z}_2\rangle$ state~\cite{wenwei18TDVPscar}. 
This state is a period-2 density wave, which can be described using a 2-atom unit cell.  The variational MPS state is parametrized by two angles, $\theta_{o}$ and $\theta_e$ that describe the state of Rydberg atoms on odd and even sites, respectively. In particular, points $(\theta_{o},\theta_e)=(0,\pi)$ and $(\theta_{o},\theta_e)=(\pi,0)$ correspond to two distinct period-2 density wave product states, $|\mathbb{Z}_2\rangle = |{\bullet}{\circ}{\bullet}{\circ}\ldots\rangle$ and its 
partner shifted by one lattice site, $|{\mathbb{Z}}_2'\rangle = |{\circ}{\bullet}{\circ}{\bullet}\ldots\rangle$.  
By projecting quantum dynamics onto this manifold using TDVP, one obtains a system of classical non-linear equations which govern the evolution of the angles, $\partial_t \theta_{o,e} = f_{a,b}(\theta_{o},\theta_e)$.  The resulting phase portrait in  Fig.~\ref{Fig:Ry}{\bf a} reveals a periodic trajectory which passes through  $|\mathbb{Z}_2\rangle$ and $|\mathbb{Z}_2'\rangle$ states, thereby explaining the revivals in the PXP model from a period-2 density wave state. 
The projection of quantum dynamics is not exact,  however, and the so-called quantum leakage, $\gamma^2$ (see Box~2), quantifies the discrepancy between quantum and variational dynamics. The leakage, shown by color scale in Fig.~\ref{Fig:Ry}{\bf a}, remains low in the vicinity of the classical trajectory, thus justifying the existence of revivals in quantum dynamics. 

Finding the classical trajectory behind quantum revivals from $|\mathbb{Z}_2\rangle$ state provided a complementary picture behind weak ergodicity breaking by eigenstates of the PXP model. This variational approach was also extended to capture quantum dynamics for PXP model from more complicated initial states or in higher dimensions~\cite{Michailidis2020}. In particular, an MPS variational ansatz with three angles $\theta_{a}$, $a=1,2,3$, was used to study dynamics from period-3 density wave state, $|\mathbb{Z}_3\rangle = \ket{{\bullet}{\circ}{\circ}{\bullet}{\circ}{\circ}\ldots}$.  TDVP projection of quantum dynamics onto such an MPS manifold gave rise to a three-dimensional dynamical system, whose Poincar\'e sections are shown in Fig.~\ref{Fig:Ry}{\bf b}. A prominent feature of this dynamical system was the existence of \emph{mixed phase space} -- the coexistence of classical chaotic regions with regions of stable motion, the so-called Kolmogorov-Arnold-Moser (KAM) tori~\cite{ArnoldBook}. The KAM tori host stable periodic orbits in their centers (one such orbit is shown by the star in Fig.~\ref{Fig:Ry}{\bf b}). This provides a practical way of finding the most non-ergodic initial state with period-3 translation invariance.  Furthermore, one may expect that the robustness of classical KAM tori to weak deformations may be used to infer the stability of the corresponding quantum model to local perturbations. The mapping of quantum dynamics onto a classical dynamical system via TDVP suggests an intriguing direction for extending the KAM theorem to quantum systems, in a way that complements other recent attempts based on weakly-broken quantum integrability~\cite{GlimmersKAM}. Pursuing this direction, however, requires a better understanding of quantum leakage, which plays a decisive role in discriminating which trajectories give rise to slow thermalization in exact quantum dynamics and which do not.

The existence of classical periodic trajectories underlying quantum revivals in the PXP model calls for exploration of semiclassical dynamics in other scarred models discussed in Section~\ref{Sec:ETH}. While the projector-based embedding scenario (iii) is, in general, unlikely to have  underlying periodic trajectories, for models with the spectrum generating algebra (i) one could imagine constructing a variational subspace using properties of $Q^\dagger$-operator.  Such a construction of classical periodic trajectories in a broader family of models could be used as a finer classification scheme for models displaying weak ETH violation in their eigenstate properties.

\definecolor{shadecolor}{rgb}{0.8,0.8,0.8}
\begin{shaded}
\noindent{\bf Box 2 $|$ Variational method for  quantum many-body scars }
\end{shaded}
\vspace{-9mm}
\definecolor{shadecolor}{rgb}{0.9,0.9,0.9}
\begin{shaded}
Optimal projection of quantum dynamics onto a given variational manifold $\mathcal{M}$, parametrized as $|\psi(z)\rangle$, can be achieved by the time-dependent variational principle (TDVP). The variational principle determines the best direction of evolution within the manifold, $\dot z \partial_z |\psi\rangle$, such that the difference between exact and projected dynamics is minimal, $\gamma^2 = \left|z \partial_z |\psi\rangle + i H |\psi\rangle\right|^2$. In order to capture quantum many-body dynamics, the manifold can be parametrized by matrix product states (MPS), whose bond dimension controls the amount of entanglement generated during time evolution~\cite{Haegeman}. Semiclassical dynamics emerges naturally in this approach when the bond dimension of the MPS is kept small, allowing to systematically study corrections to the mean-field behavior. 

While the TDVP approach can be applied to any quantum system, its application to the PXP model with MPS of bond dimension 2 results in a particularly elegant and analytically-tractable classical dynamical system~\cite{wenwei18TDVPscar}.  In order to describe coherent many-body oscillations in the PXP model,  inspired by the mean-field picture, one may attempt to use the manifold of product states,  $\ldots (\cos\theta_e|{\circ}\rangle+\sin\theta_e |{\bullet}\rangle)(\cos\theta_o|{\circ}\rangle+\sin\theta_o |{\bullet}\rangle)\ldots$, parametrized by two degrees of freedom -- the angles, $\theta_e$ and $\theta_o$. These angles describe the state of atoms on even and odd sites in the chain.  However, such a state violates the Rydberg blockade condition. Instead, the Rydberg constraint can be satisfied in the manifold of matrix product states, where one assigns the tensor $A^{{\bullet}} = \sigma^+$ to the $|{\bullet}\rangle$ state. Property $A^{{\bullet}}\cdot A^{{\bullet}} = (\sigma^+)^2=0$ of Pauli raising operators $\sigma^+ = (\sigma^x+i\sigma^y)/2$ guarantees that the weight of configurations with any two adjacent states $|{\bullet}{\bullet}\rangle$ vanishes. Application of TDVP to such states results in the system of classical non-linear differential equations for $\theta_{e,o}$.   The projection of quantum dynamics is not exact,  however, and the so-called quantum leakage, defined by $\gamma^2$ above, quantifies the instantaneous disagreement,  see Fig.~B2. 

The mapping of quantum dynamics onto a classical dynamical systems provides a useful method of studying many-body systems without an obvious semiclassical limit.  Finding suitable manifolds and extending this approach to dynamics in other models that have non-thermal eigenstates remains an outstanding open problem. 

\begin{center}
\includegraphics[width=0.9\columnwidth]{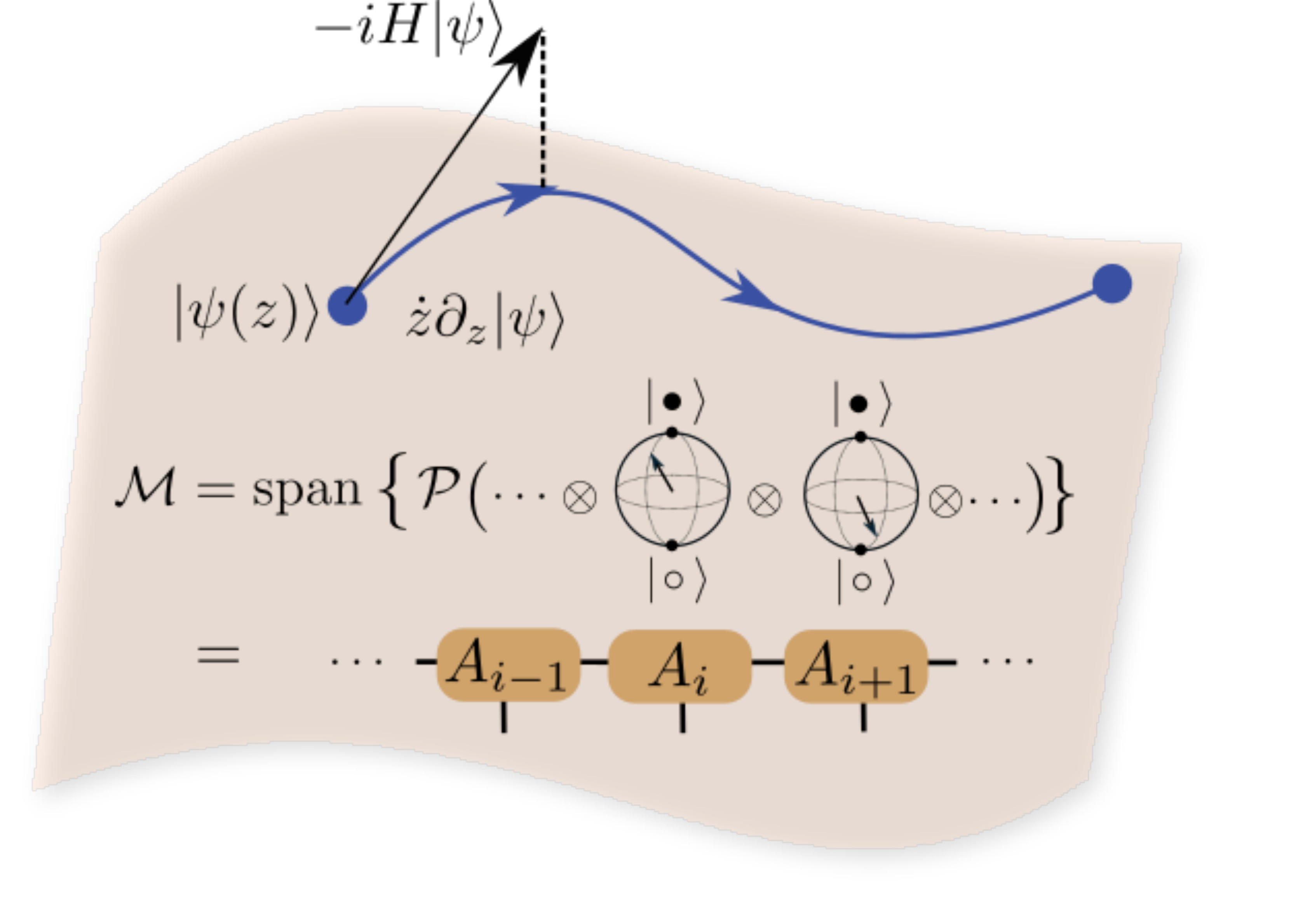}
\end{center}
\small {\noindent{\bf Figure B2 $|$ Variational method for quantum many-body scars.} 
Time-dependent variational principle captures the optimal projection of quantum dynamics onto a given variational manifold. The studies of PXP model use matrix product states (MPS) as a variational manifold, where each atom is described by a single angle. The presence of the Rydberg blockade, implemented by projector $\cal P$, results in non-zero entanglement, which requires a bond dimension 2 of MPS tensors, $A_i$.  
 }
\end{shaded}

\section{Perspective}\label{Sec:5}

Recent studies of quantum many-body scars and related phenomena, described in this article, have revealed a new kind of dynamical behavior in many-body systems -- weak ergodicity breaking. The defining feature of weak ergodicity breaking is strong dependence of relaxation dynamics on the system's initial configuration. In contrast to conventional ergodic systems, certain many-body states are long-lived, exhibiting parametrically slow relaxation, as opposed to other initial configurations that thermalize quickly. These findings have pointed out that ETH, in its strong form, does not apply to a large class of systems that host anomalous, non-thermal eigenstates, and exhibit long-time coherent dynamics. 

The discovery of quantum many-body scars serves as a reminder that the understanding of thermalization and chaos in quantum many-body systems is far from complete and requires building a more elaborate theory and the accompanying methodological toolbox. The eigenstate description of quantum many-body scars revealed a common pattern in different families of scarred systems -- an emergent decoupled subspace within the full many-body Hilbert space. While we presented a few mechanisms that could underlie the existence of such a non-thermal subspace, their complete classification is currently lacking. In particular, more work is needed to understand the precise connection between many-body scarring and two broad classes of weak ergodicity breaking phenomena: theories with confinement~\cite{Calabrese16, Konik2, Yang2020, CastroAlvaredo2020} and lattice gauge theories~\cite{, Magnifico2020, Chanda2020, Borla2020}. 
The latter, in particular the one-dimensional quantum link model, which has recently been realized in a Bose-Hubbard quantum simulator~\cite{Yang2020QLM}, have intriguing connections with the PXP model~\cite{Surace2019}, and it would be interesting to explore possible connections  in higher dimensions~\cite{Celi2020}. If accomplished, the classification of non-thermal subspaces could shed light on the physical mechanisms that lead to quantum many-body scars.  Empirically, kinetically-constrained systems appear to be especially likely to host quantum many-body scars, hence it would be important to understand the role of dynamical constraints which can lead to glassy-like dynamics~\cite{Pancotti2019,Roy2020,
Lan2017_2,Feldmeier2019, Hart2020}.

More generally, the variational approach to scarred dynamics complements recent efforts in understanding parallels between classical chaos measures, on the one hand, and thermalizing quantum dynamics and its underlying transport coefficients on the other hand~\cite{Leviatan2017,Hallam2019}.  While many-body scarred models are quantum chaotic, their properties deviate from other chaotic models such as the Sachdev-Ye-Kitaev (SYK) model~\cite{Sachdev1993,Kitaev2015}, which has attracted much attention as the fastest scrambler of quantum information~\cite{Maldacena2016}. Such deviations could be identified and studied more systematically using the variational approach and its resulting mixed phase portraits, which appear to be a generic feature of local Hamiltonians. Armed with a better understanding of quantum leakage from the variational manifold, this method could be used to detect atypical behavior and absence of maximal scrambling in general models that do not have an obvious large-$N$ limit. One of the ultimate future challenges is to bring together this approach to real-time dynamics with approaches targeting directly the properties of many-body eigenstates. This challenge may be attacked  using quantum information techniques, which, for example, allow one to link revivals and eigenstates properties~\cite{Alhambra2019}. Related techniques, as well those used to bound relaxation times from below in prethermal system, may give crucial insights into the response of QMBS to perturbations, and their dynamics at long times.

Finally, there is strong experimental and practical interest in quantum many-body scars. Many-body scarred revivals provide a mechanism for maintaining coherence, despite the presence of interactions which normally scramble local quantum information. In particular, scars in Rydberg chains  --  so far the main experimental realization of the phenomenon -- have already been utilized for the preparation of specific entangled states~\cite{Omran570}. This application made use of the quantum control based on the variational TDVP approach and its identification of entangled periodic trajectories that simultaneously have small quantum leakage. This suggests that scars may have wider range of applications, for example in protected state transfer on quantum networks or in quantum sensing. Such applications require deeper theoretical understanding of the effects that protect the coherence of scars, as well as the development of  general experimental techniques for creating them on demand, e.g., using pumping protocols in dipolar Bose gases~\cite{Kao2020} or by tilting the Fermi-Hubbard model~\cite{Scherg2020}. Looking back at the field of  single-particle scars, which have been realized in a multitude of experimental settings, one may also expect many-body quantum scars to be relevant to a much broader family of quantum simulators beyond Rydberg atoms and perhaps even in solid state materials. In all these platforms, quantum many-body scars may serve as a vehicle for controlling and manipulating many-body states, which may prove useful in a range of quantum technology applications. At this stage, the new field of weakly broken ergodicity is still at its infancy, and experimental progress will undoubtedly yield more surprises in the near future.

\section{Acknowledgements}

We are grateful to our collaborators Kieran Bull, Soonwon Choi, Jean-Yves Desaules, Wen Wei Ho, Ana Hudomal, Misha Lukin, Ivar Martin,  Hannes Pichler, Nicolas Regnault, Ivana Vasi\'c,  and in particular Alexios Michailidis and Christopher Turner, without whom this work would not have been possible. Moreover, we benefited from useful discussions with Ehud Altman, B. Andrei Bernevig, Anushya Chandran, Paul Fendley, Vedika Khemani, and Lesik Motrunich. 
M.S. was supported by European Research Council (ERC) under the European Union's Horizon 2020 research and innovation program (Grant Agreement No. 850899). D.A. was supported by the Swiss National Science Foundation and by the European Research Council (ERC) under the European Union's Horizon 2020 research and innovation program. Z.P. acknowledges support by the Leverhulme Trust Research Leadership Award RL-2019-015.

\end{document}